\documentclass[sigconf,natbib=false]{acmart}
\usepackage{booktabs}
\AtBeginDocument{%
  }

\setcopyright{acmlicensed}
\copyrightyear{2025}
\acmYear{2025}
\acmDOI{XXXXXXX.XXXXXXX}
\acmConference[EASE 2025]{The 29th International Conference on Evaluation and Assessment in Software Engineering}
\acmISBN{978-1-4503-XXXX-X/2025/XX}

\usepackage{algorithm2e}
\usepackage{graphicx}
\usepackage{textcomp}
\usepackage{nicematrix}
\usepackage{colortbl}

\newcommand*{\mybox}[2]{\newline\newline\noindent\colorbox{#1!30}{\parbox{.98\columnwidth}{#2}}\newline}

\RequirePackage[
  datamodel=acmdatamodel,
  style=acmnumeric,
  ]{biblatex}

\begin{document}

\title{An Empirical Study on Common Defects in Modern Web Browsers Using Knowledge Embedding in GPT-4o}

\author{Rahul Singh}
\affiliation{%
  \institution{Gannon University}
  \city{Erie}
  \state{PA}
  \country{USA}
}
\email{singh041@gannon.edu}

\author{Yousuf Sultan}
\affiliation{%
  \institution{Gannon University}
  \city{Erie}
  \state{PA}
  \country{USA}
}
\email{mir002@gannon.edu}

\author{Tajmilur Rahman}
\affiliation{%
  \institution{Gannon University}
  \city{Erie}
  \state{PA}
  \country{USA}
}
\email{rahman007@gannon.edu}

\author{Sri Vidya Puttareddygari}
\affiliation{%
  \institution{Gannon University}
  \city{Erie}
  \state{PA}
  \country{USA}
}
\email{puttared001@gannon.edu}

\begin{abstract}
Technology is advancing at an unprecedented pace. 
With the advent of cutting-edge technologies, 
keeping up with rapid changes are becoming increasingly challenging.
In addition to that, increasing dependencies on the cloud technologies have imposed enormous pressure on modern web browsers leading to adapting new technologies faster and making them more susceptible to defects/bugs.
Although, many studies have explored browser bugs, a comparative study among the modern browsers generalizing the bug categories and their nature was still lacking. 
To fill this gap, we undertook an empirical investigation aimed at gaining insights into the prevalent bugs in Google Chromium and Mozilla Firefox as the representatives of modern web browsers.
We used GPT-4.o to identify the defect (bugs) categories and analyze the clusters of the most commonly appeared bugs in the two prominent web browsers.
Additionally, we compared our LLM based bug categorization with the traditional NLP based approach using TF-IDF and K-Means clustering.
We found that although Google Chromium and Firefox have evolved together since almost around the same time (2006-2008), Firefox suffers from high number of bugs having extremely high defect-prone components compared to Chromium. 
This exploratory study offers valuable insights on the browser bugs and defect-prone components to the developers, enabling them to craft web browsers and web-applications with enhanced resilience and reduced errors.
\end{abstract}




\maketitle

\section{Introduction}
Modern technology is advancing rapidly, with innovations like 3D gaming, high-details video games, virtual reality, 3G to 6G network advancements, improved network, and high-speed cloud computing. 
Software applications that rely on advanced technologies must evolve to keep up with the rapidly changing demands. 
Web browsers have become essential due to the growing need for portability and internet access. 
To adapt with this, modern web browsers are evolving faster than ever. 
However, this lightning speed has also led browsers to increased bug reopening rates (7\% increase of bugs in Firefox~\cite{souza2014rapid}).

% Bug reports of Chromium and Firefox
Both Google Chromium and Firefox use publicly available bug tracking systems (Chromium Issue Tracker~\cite{chromiumbug} and Bugzilla~\cite{bugzilla}). 
% A little background on Chromium and Firefox
Firefox and Chromium both have a similar lifespan starting their first version release in 2006~\cite{firefoxrelease} and 2008~\cite{chromerelease} respectively.
%Although Firefox was officially launched in 2004, its code-base has a rich and extensive history that can be traced back to Netscape, which was the second influential web browser after Mosaic~\cite{baysal2011tale}.
Both browsers are leading modern web browsers that have continuously evolved, competed, and embraced new technologies over time.
They are two of the most popular open source browsers~\cite{popularBrowsers} and have a very similar architecture that complies with the standard reference architecture of modern web browsers proposed by Rahman et al.~\cite{rahman2019modular}. 
%Both share a similar development strategy and release structure, exhibiting a strong resemblance to each other. 
%After their initial release, both Firefox and Google Chromium adopted a 6-week release model. 
%Later on, they both increased their release frequency around the same time. 
%Firefox moved to a 4-weeks cycle in late 2019~\cite{firefoxrelease}, and Chromium followed in 2021~\cite{chromium4-weeks}.

% Why study common defects 
Despite numerous studies that have been carried out to analyze, predict, and classify software bugs from different perspectives, very few research~\cite{baysal2011tale, song2022r2z2, zaman2011security, rocha2016characterizing, lal2012comparison} have been conducted specifically on bugs in web browsers.
Due to the enduring popularity and successful adaptation to technological advancements of Mozilla Firefox and Google Chrome, we have been motivated to conduct a comparative analysis among these web browsers to explore the common defects, bug categories, and how it affects the browser components from a unique perspective of using Large Language Model.

\textit{\textbf{Novelty:}} -- 
Although there are studies focused on categorizing bugs~\cite{rocha2016characterizing} and investigating specific types of browser-related bugs~\cite{zaman2011security}, no empirical research has yet explored bugs in modern web browsers from a broader, generic perspective to identify their commonalities and differences.

\textit{\textbf{Contribution:}} -- 
In this study, we have investigated the common bugs in Firefox and Chrome as the representatives of modern web browsers, and explored how bugs are affecting browser components, and developers’ effort. 
We embedded contextual knowledge from the browser documentations, commit messages, and bug descriptions into pre-trained GPT-4.o to analyze the nature of the bugs in model web browsers, and grouped them into different categories. Additionally, we categorized bugs using traditional NLP technique using TF-IDF and compared the performance with the LLM based approach.

We believe our findings will benefit software engineering community especially the modern web-browser developers to understand the defect-prone components, effort-demanding areas in modern browsers.
This will help developers take informed decisions to prioritize tasks, or take proactive actions with caution before modifying certain areas in a web-browser.
This empirical analysis is based on the following research questions:\\

\textbf{RQ1: How does LLM identify common bugs in modern web browsers? How does NLP perform in contrast?}\\
Our primary goal is to identify the common defects in modern web browsers using LLM. Since LLMs are built on the foundation of NLP and Deep Learning. 
Answer to this question will also help us understand whether the fundamental NLP techniques are still better choice for categorizing the bugs from the bug descriptions. 

\textbf{RQ2: What are the common defects in modern web browsers i.e., Firefox and Chromium?}\\
We wanted to explore the categories of bugs that appear frequently in both browsers, along with their occurrence patterns. 
This will allow developers to strategically allocate resources to rectify the most prevalent bugs to enhance the general quality and dependability of modern web browsers.

\textbf{RQ3: Do highly effort-consuming bugs exist, and if so, are they prevalent in both browsers?}\\
Bugs require effort to attempt a fix. Often certain bugs demand enormous effort from the developers. 
We wanted to understand if there are certain types of bugs that are highly effort consuming and if they are common in both Firefox or Chromium. 
This will benefit both researchers and industry experts finding a long-term solution for preventing these bugs.

\textbf{RQ4: What are the common components in modern web browsers, and which of them need more attention from developers?}\\
Firefox and Chromium began development at a similar point in time and have evolved alongside each other. 
Both browsers share comparable components, which can lead to similar types of bugs. 
Answer to this question will examine how defect-prone specific browser components are, helping developers take necessary precautions.

% Rest of the paper
The rest of the paper is organized as follows. 
Section~\ref{background} discusses the existing research works related to this study and compares with our approach and goals. 
Section~\ref{data} explains the data and the data collection process, Section~\ref{method} describes the study method, 
Section~\ref{result} discusses the findings and answers to the research questions followed by Section~\ref{threats} which discusses the threats to validity. 
Finally, Section~\ref{conclusion} discusses the paper and concludes with our future plans.

\textbf{Replication Package:} To comply with the open science policy we made our study resources publicly and anonymously available\footnote{Replication package: https://figshare.com/s/a99e9053108bf300e579}. 
The replication package contains the datasets, figures, and Python scripts.

\section{Background and Related Works}
\label{background}
Software bug/defect is one of the most extensively studied areas~\cite{mutasim2023leveraging}\cite{hammouri2018software}\cite{tran2020analysis}\cite{monperrus2018automatic}\cite{cairo2018impact}\cite{rodriguez2020bugs}\cite{cotroneo2019bad} in the literature.
Numerous studies have been conducted to empirically analyze~\cite{cotroneo2019bad}\cite{tran2020analysis}, predict~\cite{hammouri2018software, immaculate2019software, khan2020hyper}, and auto-repair~\cite{monperrus2018automatic, allamanis2021self} bugs. 
Bug reports have also been studied in web browsers~\cite{medeiros2020improving}\cite{song2022r2z2} to localize bugs~\cite{wu2018changelocator}\cite{pradel2020scaffle} from the crash-reports, to prioritize test-cases~\cite{feng2015test}, and analyzing flaky tests~\cite{rahman2018impact}.

%Browser related studies
% We found the exploratory study conducted by Baysal et al.~\cite{baysal2011tale} on Mozilla Firefox and Google Chrome web browsers highly pertinent to our research where Baysal et al. empirically analyzed the two web browsers and compared on several factors. 
% Baysal investigated and compared the release cycle lengths, release pattern, defect-proneness, bug-fix speed, user adoption trend, and reliability. 
% In contrary, our study is primarily focused on bugs, their categories in relation to the components and developers' effort.

%consise
Baysal et al.~\cite{baysal2011tale} conducted an empirical analysis of Mozilla Firefox and Google Chrome, comparing release cycle lengths, patterns, defect-proneness, bug-fix speed, user adoption trends, and reliability. While their study provides valuable insights into these factors, our research focuses specifically on bug categorization, their association with components, and the developers' effort in addressing them.

% Baysal et al.~\cite{baysal2011tale} found that Firefox was experiencing higher number of bugs compared to Chrome until their seventh release version.
% This is because Firefox introduced a lot of new code that time and got a huge spike of 24\% increase in bugs although the ``severe'' bugs did not increase as much.
% It has been more than twelve years since Baysal et al. empirically studied the two giant open source web browsers of two decades. 
% However, we haven't found any empirical study investigating bugs, defect patterns, or other issues specifically focused on bugs among the modern web browsers since then. 

%consise
Baysal et al. [8] observed that Firefox experienced more bugs than Chrome up to its seventh release, driven by a 24\% spike in bugs due to the introduction of significant new code, though 'severe' bugs increased less dramatically. Despite being over twelve years old, their study remains one of the few empirical analyses of Firefox and Chrome. Since then, no comprehensive study has specifically examined bugs, defect patterns, or related issues in modern web browsers.

% A comparative study of bugs between iOS and Android operating systems has been conducted by Aljedaan et al.~\cite{aljedaani2019comparison}.
% Although, they have done the comparison based on the Firefox and Chromium bugs in iOS and Android, this is basically a comparison between the two operating systems. 
% In this study the authors wanted to see how defect-prone are the two browsers in iOS platform compared to Android.
% Furthermore, Aljedaan et al. examined three aspects of analysis which were the rate of submitting bug reports, the duration of bug resolution, and the classification of bugs using topic modeling techniques to categorize them. 
% They discovered notable variations between these factors and concluded that the number of bugs in the Android version of the browsers was higher than in the iOS version.

%consise

Aljedaan et al. [28] conducted a comparative study of Firefox and Chromium bugs on iOS and Android, essentially contrasting the two operating systems. They analyzed the rate of bug report submissions, bug resolution times, and bug classification using topic modeling. Their findings revealed significant differences across these aspects, concluding that the Android versions of the browsers exhibited a higher number of bugs than their iOS counterparts.

Zaman et al.~\cite{zaman2011security} conducted an empirical study that explored the differences in bug-fix characteristics across security, performance, and other types of bugs. Their work examined key metrics such as the number of lines changed, the number of files affected, and the entropy to measure the complexity of bug a fixe. They found that security bugs were generally more complex to address than performance or other types of bugs, requiring more developers with greater experience, yet they are often resolved more quickly. Zaman's study primarily focused on the Firefox web browser only. It provides valuable insights into how different types of bugs are resolved in Firefox.
In contrast, our research offers a more comprehensive analysis of the natures of bugs in modern web browsers. We expand the scope by including additional bug types and investigating how these bugs affect the browser components in general. By conducting the study in both Firefox and Chrome, our study aims to build on and generalize the findings, contributing to a deeper understanding of bug categories and their influence across modern web browsers.

Rocha et al.~\cite{rocha2016characterizing} studied bug resolution workflows in multiple open source projects.
Rocha's study uses Bug Flow Graphs (BFGs) to map out the paths bugs take through the resolution process.
On the other hand, our research takes a fresh approach by using GPT-4's knowledge embedding technique to study common defects in modern web browsers.

Lal and Sureka~\cite{lal2012comparison} analyzed seven bug categories in Google Chrome. They primarily focused in Google Chrome's bug reports to explore metrics such as Mean Time to Repair (MTTR) and Debugging Efficiency.

Recent studies related to browsers include empirical studies on browser threats~\cite{lin2020fill}, performance~\cite{jin2024impact}\cite{bogdan2024empirical}, security~\cite{zaman2011security} and privacy~\cite{akhavani2021browserprint}\cite{zafar2023comparative}.

Lin, Ilia, and Polakis \cite{lin2020fill} investigated browser autofill vulnerabilities which allow hidden fields to exfiltrate sensitive user data, affecting 5.8\% of Chrome and 24.5\% of Firefox forms. To improve security, they proposed a Chrome extension that detects and blocks risks and share the findings with vendors.

%According to Jin, Li, and Zou \cite{jin2024impact}, certain Chrome extensions increase load times and energy use, suggesting optimized design and usage settings. Andrei and Malavolta \cite{bogdan2024empirical} showed that CSS prefix slows down mobile and web applications without a noticeable energy impact, advising removal to improve speed. Both studies underscore the performance trade-offs of third-party Web tools.

% TF-IDF is better over word2vec models
TF-IDF is a widely adopted technique in NLP for text service-processing. 
TF-IDF identifies the topics of a document~\cite{tandel2016multi} and is one of the most widely used techniques in NLP~\cite{el2021automatic}.
Khan et al.~\cite{khan2019extractive} used TF-IDF and K-Means clustering to summarize texts from input documents collected from various news articles.
Pimpalkar et al.~\cite{pimpalkar2020influence} studied the influence of pre-processing strategies on the performance of machine learning models where they used Bag of Words (BoW), TF-IDF as part of the ``service-processing'' technique.

%Application of machine learning 
Many studies on software bugs have used Machine Learning Techniques (MLT) for predicting~\cite{pisolkar2022empirical}\cite{hammouri2018software}\cite{immaculate2019software}, classifying~\cite{khan2020hyper}\cite{ramani2012}, and different types of analysis. 
For the automated classification and triaging of bug reports MLT have been used extensively in a number of research works~\cite{di2002approach}\cite{anvik2006automating}\cite{bhattacharya2010fine}~\cite{podgurski2003automated}.
NLP and clustering techniques together are also popular for analyzing texts found in bug-reports, commit messages, bug descriptions etc~\cite{limsettho2016unsupervised}\cite{limsettho2014automatic}. 

Limsettho et al.~\cite{limsettho2016unsupervised} developed an automated framework to categorize bug reports, according to their grammatical structure without the requiring labeled data. 
Limsettho and his colleagues utilized topic modeling and clustering algorithm to group bug reports based on their textual similarity. 
%Their framework allows customization using a modular approach, and they developed a new clustering labeling algorithm to assign labels to each group of bug reports in the topic space. 
%Their approach involves NLP-chunking to generate representative labels for each cluster. 

Meng et al.~\cite{meng2022automatic} proposed an automatic bug report classification method incorporating both text information and the intention behind reports, achieving significant accuracy improvements using a combination of BERT and TF-IDF for feature extraction. 
Their focus was on binary bug categorization (bug vs. non-bug) across multiple ecosystems using five machine learning classifiers. 
In contrast, our study expands beyond binary classification by examining a broader range of bug categories (e.g., performance, usability, security) and their relationships with defect-prone components in web browsers. 
%Additionally, while Meng et al. utilized intention labels for improved classification, our approach leverages advanced Large Language Models (LLMs) for richer semantic insights, enabling automated and context-aware categorization. These enhancements provide more actionable insights for developers and contribute to a deeper understanding of bug patterns

Although many studies have been conducted to detect, predict, categorize bugs in different settings using Large Language Models (LLM), as of today, we haven't found any empirical study comparing the modern web browsers to explore the similarities and dissimilarities among the modern browsers from a generic perspective. 
%from bug reports and revision history exploring the natures of browser bugs and the influence on browser components and developers' effort. 
%In this study we were interested in understanding the types of bugs that are prevalent in modern web browsers. 
%We used LLM and compared its performs over traditional NLP based approach. 

\begin{table}[htbp]
\caption{Chromium Components}
\begin{center}
\label{tab:ch-components}
\begin{tabular}{|p{1.7cm}|p{1.7cm}|p{1.7cm}|p{1.7cm}|}
\hline
\multicolumn{4}{|c|}{\textbf{Component Names}} \\
\hline
Android & App & Ash & API\\
\hline
Athena & Aura & Autocomplete & Autofill \\
\hline
Blink & Bookmarks & Breakpad & Cache \\
\hline
CC & ChromeOS & Cloud Print & Content \\
\hline
Cookies & \textbf{Courgette} & Crypto & CSS \\
\hline
Design & Desktop & Device & DNS \\
\hline
Docs & Dom & Download & Enterprise \\
\hline
Extensions & Fuchsia & GFX & Google APIs \\
\hline
Google Update & GPU & Headless & Infra \\
\hline
Internals & IO & iOS & IPC \\
\hline
Keyboard & Layout & Mandoline & Mash \\
\hline
Media & Mobile & Mojo & Monitor \\
\hline
Native Client & NET & Notification & O3D \\
\hline
OS & Parser & Password & Performance \\
\hline
Permission & Platform & PPAPI & Preferences \\
\hline
Printing & Privacy & Profile & Remoting \\
\hline
Reporting & RLZ & Sandbox & SDCH \\
\hline
Search & Security & Services & Shell \\
\hline
Skia & Speed & Storage & Sync \\
\hline
Themes & Toolbar & Tools & UI \\
\hline
Upboarding & URL & Utils & V8 \\
\hline
Views & Webapps & Weblayer & Webrunner \\
\hline
Webstore & Webtools & Widget & Webkit\\
\hline
\end{tabular}
\end{center}
\end{table}

\section{Data}
\label{data}
Since 2008, Chromium has evolved significantly~\cite{chromeblog}, and our study covers bug reports for all 122 versions starting from 5.0 until 127.0. 
Chromium boasts a total of 131 release versions as of today.
%with their publicly accessible bug reports~\cite{monorailbug}.
%Firefox
Mozilla Firefox~\cite{firefoxrelease} has maintained a release cycle comparable to Chromium resulting in a similar number of annual releases. 
Our study encompasses the entire commit history of Firefox since 1998. 
In this section, we describe how we collected commit history and bug records from four distinct sources.

\subsection{Repository Mining}
\label{commitdata}
We obtained the development history by cloning Chromium and Firefox's GitHub mirrors~\cite{gitchromium, gitfirefox} and then stored the data into Postgres database with the help of Perl scripts following an approach practiced by Bird et al~\cite{bird2009}.

\subsection{Commits and Revision Details}
We extracted nearly 937K, and 1493K commits with revision data from Firefox and Chromium repositories respectively. 
%The git revision data for each commit gives us the details  on which files were modified, how many likes added/deleted, if the file was renamed, etc. 
In Git, every commit is identified by a unique ``commit hash'' and contains essential information like commit message, author (developer), date when the code change was committed, changes, modified files, release tags, repository heads, and more. 
We structured these details into distinct database tables to enable efficient SQL queries for extracting attributes and metrics to find answers to our research questions.

\subsection{Bug Reports}
The bug reports include priority, type, status, and, most importantly, the bug description summaries. 
The summary text where crucial for us to apply NLP techniques and knowledge embedding in LLM. 
We stored the bug reports for both browsers into our Postgres database. 
\begin{table}[htbp]
\caption{Firefox Components}
\begin{center}
\label{tab:fx-components}
\begin{tabular}{|p{1.7cm}|p{1.5cm}|p{1.5cm}|p{1.7cm}|}
\hline
\multicolumn{4}{|c|}{\textbf{Component Names}} \\
\hline
About & Access & Accessible & Addons \\ 
\hline
Addressbook & Android & Api & Application \\
\hline
Autocomplete & b2g & Bookmarks & Branding \\
\hline
Build & Cache & Calendar & Camino \\
\hline
Canvas & Caps & CCK & CDP \\
\hline
Chimera & Cmd & Code-quality & Config \\
\hline
Content & Cookies & CSS & Desktop \\
\hline
Devtools & DNS & Docshell & Dom \\
\hline
Download Manager & Editor & Embedding & Enterprise Policies \\
\hline
Expat & Experiments & Extensions & Formautofill \\
\hline
 Freetype2 & GC & GFX & Glean \\
\hline
GPU & Grendel & HAL & ICU \\
\hline 
Infrastructure & IPC & Keyboard & Layout \\
\hline
Mailnews & Media & Memory & Message \\
\hline
Mobile & Modules & Mozbuild & Mozglue \\
\hline
Mozperftest & Net-Monitor & Network & Notification \\
\hline
NSPR & OS & Parser & Password \\
\hline
Performance & Permission & Plugin & Preferences \\
\hline
Printing & Privacy & Profile & RDF \\
\hline
Remote & Reporting & Screenshots & Search \\
\hline
Security & Services & Servo & Shell \\
\hline
Signaling & Src & Storage & Sync \\
\hline
Taskcluster & Themes & Third-Party & Toolbar \\
\hline
 Toolkit & Tools & UI & Uriloader \\
\hline
Views & Wallet & Webapps & Web Platform \\
\hline
Webservice & Webtools & Widget & XPCOM \\
\hline
XPFC & XPFE && \\
\hline
\end{tabular}
\end{center}
\end{table}

\subsection{Components}
In the Chromium project each directory contains a DIR\_METADATA file that designates itself to a component. 
%For example, a DIR\_METADATA file is considered to belong to the same component as their parent directory. 
Components are structured in a format of either ``COMPONENT'' or ``COMPONENT$>$SUB-COMPONENT''. 
By mining the DIR\_MATADATA files we initially identified 856 and 453 components including sub-components for Chromium and Firefox respectively.
We further cross-verified these component names with Chromium's documentation~\cite{chromiumComponents} and refined them based on the following elimination criteria.

\begin{enumerate}
    \item Not a component according to documentation.
    \item Contains utility / helper libraries.
    \item Contains test code.
    \item Contains binaries/executable only.
    \item Contains local executors.
    \item Contains third-party source code.
\end{enumerate}

For example, we excluded ``Gin'' from our initial list as it serves as a utility for V8 rather than a component~\cite{gin-not-a-component}. 
Conversely, ``Courgette'' was added as a component since it manages micro-updates~\cite{courgette-is-a-component} which is crucial for browser updates in modern web-browsers.
Finally, we end up with a list of 106 and 92 components for Firefox and Chromium respectively as shown in table~\ref{tab:ch-components} and~\ref{tab:fx-components}.

\begin{figure}[!ht]
  \centering
  \includegraphics[width=0.5\textwidth, height=0.5\textheight]{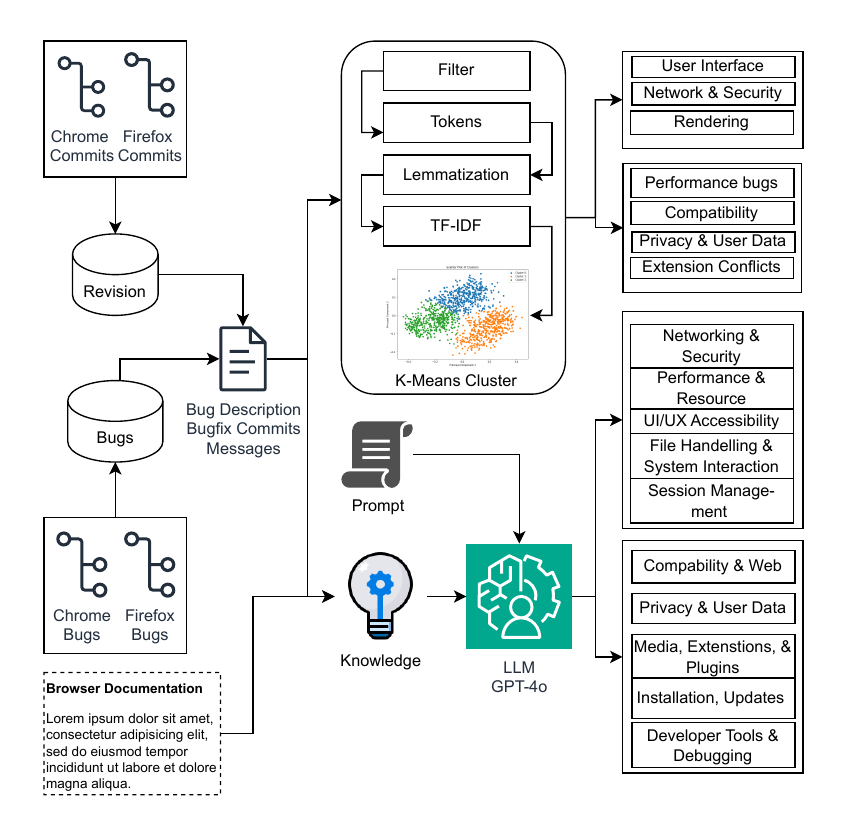}
  \caption{Categorizing Bugs in Chrome and Firefox.}
  \label{fig:methodology}
\end{figure}

\section{Methodology}
\label{method}
Once the data was collected our methodology began with a pre-processing step to synthesizes the data for analysis. 
Figure~\ref{fig:methodology} outlines how we obtained the bug categories using NLP and through knowledge embedding step by step.

\subsection{Data Pre-processing}
%To address RQ2 and RQ3, it was essential to establish connections between code repository information and bug reports. 
%We did this by extracting bug IDs from commit messages in the Git repository data. 
%This enables us to link commits to bug reports as Chromium's code-commit guidelines require bug IDs in commit messages for bug-related changes~\cite{chromium-commit-rule}.
In order to answer our third research question (RQ3), we needed to safely collect the bug descriptions and the associated component information. 
The bug reports contain component information identifying the affected component for a particular bug. 
%that tells us about which component is affected by a particular bug. 
%Although bug reporters follow a common convention while filling out component information, 
There are many occasions where one bug is affecting multiple components. 
In such cases the bug-reporter adds the names of all affected components in a chain (\textit{Ex. ``Plugin--Permission--Tools''}). 
We synthesized this into a many-to-many relational table with two columns ``Bug ID'' and ``Component Name''.

Furthermore, the component names in the bug data did not fully match the component lists we collected for the two browsers, as described in Section~\ref{data}.
For our exploratory research, we needed an organized dataset with a well maintained relationship to keep track of all entities such as bugs, components, commits, and commit-revisions.
We manually verified all component names in the bug data against the original component list and linked them to their associated bugs.

The commit data collected from the repositories contains log messages, where potential bug-fix commits provide in-depth insights into a bug, as these messages are written by developers after code investigation. 
Therefore, it was necessary to identify bug-fix commits so that we can utilize the commit messages to categorize bugs and answer RQ2. 

To address RQ2, RQ3, and RQ4, we needed to establish a sound relationship between the code repository data and the bug reports.
As the first step, this involved extracting bug IDs from commit messages in the Git revision data enabling the linkage between commits and bug reports. 
%We used Python scripts to label the code revision data with components based on file names and directory paths.
To associate the revision data with components we wrote Python scripts to label each git revision data with the relevant component based on file paths and directories. 
For example, if there is a path ``services/service manager/sandbox/mac'' we labeled it as `Services' component. 
For example, the path 'ios/chrome/browser/ui' was labeled as the 'UI' component instead of 'iOS' because the code in this path specifically relates to UI functionality.
The process of labeling the revision data with component ID was based on the following algorithm.

\begin{algorithm}[ht]
  \caption{Labeling Git Revision Data with Component}
  \KwData{$String[] paths$, $String[] components$;}
  \KwResult{Labeled $path$;}
  \For{each $path$ in $paths$}{
    $String[] dirs \gets $ directories in $path$\;
    \For{$i=dirs.length-1$ to $0$}{
      $currentDir \gets dirs[i]$\;
      \eIf{$currentDir$ is in $components$}{
        $path.label$ $\gets$ $currentDir$\;
      }{
        $currentDir \gets dirs[i--]$\;
      }
    }
  }
\end{algorithm}

The second step involved identifying bug-fix commits by retrieving bug IDs from commit messages. 
We considered the commits that contained bug IDs in the commit messages as bug-fix commits
%since developers in both Firefox and Chromium consistently followed the convention of including bug IDs in commit messages when code-changes were related to a bug. 
%The bug was actually fixed in this commit was not our concern. Because the presence of the big ID in the commit message indicates the commit was at least attempted to fix a bug. Hence, this commit is part of the effort spent for that bug.
This facilitated the correlation of bugs with code revision data.

%\subsection{Finalizing Data Preparation}
The pre-processing of Git revision data labeling with component and bug IDs marked a crucial step in preparation of the data. 
The dataset encompassed approximately 1.5M and 0.9M distinct commits from Chromium and Firefox respectively.
Table~\ref{tab:data-sum} presents a summary of the data that shows the number of commits (Comm.), churns (Ch.), bugs, bug-fix commits (BF Comm), and components (Comp.).

\begin{table}[htbp]
\caption{Data Summary}
\begin{center}
\label{tab:data-sum}
\begin{tabular}{@{}lrrrrr@{}}
\toprule
& Comm. & Ch. & Bugs & BF Comm. & Comp. \\
\midrule
Firefox     & 0.9 M & 6 M & 370K & 700K & 106 \\
Chromium    & 1.5 M & 11 M & 143K & 325K & 92 \\
\bottomrule
\end{tabular}
\end{center}
\end{table}

%Figure~\ref{fig:dev-activity-box-fx} illustrates notable differences in dev\_effort (developers' effort) and churn\_dev (churn per developer) between the two. 
%Dev\_effort is computed as $(commit + churn) / developers$.
%
%\begin{figure*}[ht]
%    \centering
%    \begin{subfigure}[b]{0.9\textwidth}
%         \centering
%         \includegraphics[width=\textwidth]{boxplot-fx-components.png}
%         \caption{Firefox}
%     \end{subfigure}
%     \hfill
%     \begin{subfigure}[b]{0.9\textwidth}
%         \centering
%         \includegraphics[width=\textwidth]{boxplot-ch-components.png}
%         \caption{Chromium}
%     \end{subfigure}
%  \caption{Developers' Activities in Modern Web Browsers}
%  \label{fig:dev-activity-box-fx}
%\end{figure*}

\subsection{Categorizing Bugs Using GPT-4.o Knowledge Embedding}

In this work, we employ large language model (LLM) to classify software bugs by embedding domain-specific knowledge into GPT-4. 
We collected browser documentation and organize them in different categories based on what information they contain. 
%The knowledge, which includes information on components and bug types, was extracted from the source documentation of Firefox and Chromium. 
%This knowledge was then used to train the model to enhance its ability to categorize and understand bugs within this specific domain.

To provide the model with rich contextual information specific to bugs and code changes, we prepared an additional set of knowledge documents from the bug descriptions in bug reports, and commit messages from Git revision logs. 
These textual sources contain valuable details about the nature and context of a bug. 
We combined the bug descriptions and corresponding bug-fix commit messages into a single dataset ``\textit{Combined Knowledge Source}'' (CKS).

\begin{figure}
  \centering
  \includegraphics[width=1\columnwidth]{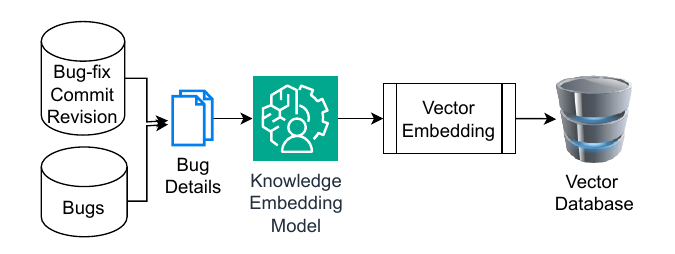}
  \caption{Knowledge Vector from Bug Reports and Commit Messages.}
  \label{fig:knowledge-vector}
\end{figure}

We utilized OpenAI's Embeddings API \textit{``text-embedding-3-small''} to transform the CKS into high-dimensional vector representations as shown in Figure~\ref{fig:knowledge-vector}. 
These vectors capture intricate relationships and subtle patterns in the bug descriptions and logs, contextualized for the domain of web browsers. 
By leveraging these embeddings, the model was able to discern complex connections between the bugs.
These vectors capture subtle details and patterns in bug descriptions and logs, allowing the model to recognize complex relationships between bugs and browser components.

%When a bug reporter reports a bug, primarily the nature of the fault and the behavior of the feature is reported. On the other hand, when a developer is assigned to work on that bug for fixing, the developer writes deeper insight into the issue at the code and component level.
%By combining these two with category definitions, we sought to capture the semantic relationships between the bug reports and predefined categories such as ''Networking \& Security``, ''Performance \& Resource Management``, and ''UI/UX \& Accessibility``. 
%We combined the bug descriptions and bug-fix commit messages as documents
%Furthermore, we analyzed the documentations for both browsers~\cite{bugDocumentationFirefox, bugDocumentationChromium} and generated a list of all possible types of bugs in browsers as shown in Table~\ref{tab:predefined-categories}.
%and call it as``\textit{Combined Knowledge Source}'' (CKS).

\begin{table}[h!]
\caption{Benchmark Categories}
\centering
\begin{tabular}{|p{2cm}|p{6cm}|}
\hline
\textbf{Category} & \textbf{Description} \\
\hline
Networking \& Security & Issues related to network connectivity, protocols, and vulnerabilities that could lead to unauthorized access or data breaches. \\
\hline
Performance \& Resource Management & Problems causing slowdowns, resource-heavy operation, or inefficiencies in system resource allocation. \\
\hline
UI/UX \& Accessibility & Bugs in the user interface or experience that affect usability, hinder navigation, or prevent the browser from being usable by individuals with disabilities. \\
\hline
Compatibility \& Web Standards & Issues arising when the browser does not function as expected across different platforms or with various web standards, including rendering errors. \\
\hline
Privacy \& User Data & Problems affecting user control over personal information, data sharing preferences, and how the browser handles data storage. \\
\hline
Media, Extensions, \& Plugins & Issues with audio/video playback, and malfunctions related to browser extensions or plugins. \\
\hline
Installation, Updates, \& User Preferences & Challenges related to installing the browser, updating it, patching software, and errors affecting the saving/restoration of user settings. \\
\hline
Developer Tools \& Debugging & Malfunctions within built-in developer tools that impede website testing, debugging, or development, including JavaScript engine bugs. \\
\hline
File Handling \& System Interaction & Issues involving browser interaction with file downloads, uploads, and local file system access. \\
\hline
Session Management \& Synchronization & Faults in how the browser manages session information and problems with synchronizing settings or data across devices. \\
\hline
\end{tabular}
\label{tab:predefined-categories}
\end{table}

%

%After embedding the contextual knowledge into the LLM we used prompts to categories the bugs. 
%\mybox{gray}{RQ1: Can we embed domain specific knowledge into LLM to classify bugs in modern web browsers? We have successfully processed ``\textit{Combined Knowledge Source}'' and embedded into \todo{GPT-4.o}}

\subsection{Categorizing Bugs Using NLP}
\label{analysis}
LLMs being pre-trained on millions of data making it capable of understanding a wide variety of context, we wanted to know how the fundamental NLP techniques perform categorizing bugs compared to the pre-trained LLM (GPT-4.o). 
%We applied cluster analysis to categorize the bugs into distinct groups. 
We begin with removing punctuations and filtering out stop words. 
As the next step we employed tokenization~\cite{keselj2009book} on the bug description text from both bug repository and commit messages. 
%Tokenization breaks the text into individual tokens or words enabling us to perform specific operations on individual words for deeper insights.
We then apply lemmatization~\cite{hapke2019natural} to reduce words to their base forms to standardized the vocabulary transforming into their base forms. 
%For instance, ``running'', ``runs'', and ``ran'' are all reduced to the lemma ``run''.
After that, we used TF-IDF to gauge word importance within the bug summary. 
TF-IDF assigned weights to the words based on its frequency and helped identifying words that are more representative of a specific bug.

We followed the cut-off policy proposed by Zhang et al.~\cite{ZHANG2020540} by balancing the precision and recall.
To improve the performance of the matrix we eliminate the words with TF-IDF score below 0.03 (irrelevant), and we ignored words having score above 0.8 (too obvious). 

To find categories we grouped common words applying K-means clustering~\cite{arthur2007k} algorithm to the lemmatized words, 'K' representing the number of clusters. 
%K-Means clustering is a well-known unsupervised machine learning algorithm, employs centroids to create clusters~\cite{arthur2007k}. 
%It assigns each data point (lemmatized words) to one of the 'K' clusters based on their similarity to the cluster centroids. 
Number of clusters were determined by the elbow curve~\cite{syakur2018integration}.% as shown in Figure~\ref{fig:elbow}.

%\begin{figure}
%    \begin{subfigure}[b]{\columnwidth}
%         \centering
%         \includegraphics[width=0.8\columnwidth]{fx-elbow-k-means.png}
%         \caption{Firefox}
%    \end{subfigure}
%    \begin{subfigure}[b]{\columnwidth}
%         \centering
%         \includegraphics[width=0.8\columnwidth]{ch-elbow-k-means.png}
%         \caption{Chromium}
%    \end{subfigure}
%  \centering
%  \caption{Elbow Curve to Determine K in K-Means Clustering.}
%  \label{fig:elbow}
%\end{figure}

We further utilized Principal Component Analysis (PCA) to reduce data dimensionality while preserving variance, to determine the optimal number of clusters~\cite{hasan2021review}. 
%PCA is a technique of dimensionality-reduction (DR) that is
%mostly used to reduce a large set of variables into a smaller one that still contains much of the details in the large set~\cite{wold1987principal}.
Finally, we identified three clusters in Firefox and two in Chromium bugs.
Table~\ref{tab:cluster-words} presents the top 15 tokens from each cluster.
We made sure to avoid any verb, adverb, or adjective to appear in the clusters.
%For example, ``wrong'', ``fails'', ``failure'', ``cause'', ``shutdown'', ``show'', these words do not indicate to any certain type of bugs since they may appear with any or almost all bug reports.
Through manual analysis of the bug descriptions containing the top tokens, we categorized the Firefox bugs as 
\textit{\textbf{C1:} UI/UX Functional Bugs}, \textit{\textbf{C2:} System Stability and Compatibility Bugs}, and \textit{\textbf{C3:} Display and Window Management Bugs}.
On the other hand, the Chromium clusters are categorized as \textit{\textbf{C1:} Cross-Platform Build \& Test Reliability Bugs}, and \textit{\textbf{C2:} System Stability and Compatibility Bugs}. 
We found C2 as the only common cluster in Firefox and Chromium. 

% as the inertia in the elbow curve significantly decreases after the second cluster.
%, as shown in Figure~\ref{fig:elbow}(b). 
% Figure~\ref{fig:scatter-pca} displays these clusters, where each dot representing a word from the tokenized bug summary. 
% This figure highlights the grouping of the most frequently appearing word tokens in the bug summary text into three categories for Firefox and two categories for Chromium.

\begin{table*}[htbp]
\caption{Bug Description Token Clusters}
\label{tab:cluster-words}
\centering
\begin{tabular}{llp{3.8cm}p{9.8cm}}
\toprule
\textbf{Browser} & \textbf{Cluster} & \textbf{Cluster Names} & \textbf{Most Frequent Tokens} \\
\midrule
Firefox   & C1 & UI/UX Functional Bugs & Page, Video, Tab, Work, Error, Text, File, Event, Content, Image, Button, Font, Test, Menu, Input \\
          & C2 & System Stability and Compatibility Bugs & Tab, Startup, Linux, Test, Report, Wayland, Window, Large, Small, Nightly, Video, Error, Process, Freeze, Client \\
          & C3 & Display and Window Management Bugs & Window, Tab, Browser, Screen, Bar, Menu, Button, Display, Monitor, Private-mode, Size, Opened, Black, Video, Title \\
\midrule
Chromium  & C1 & Cross-Platform Build and Test Reliability Bugs & Flaky-test, Mac, Linux, Window, Bot, Upstream, Platform, Web, Builder, Android, Fuchsia, Build, Release, IO, Debug \\
          & C2 & System Stability and Compatibility Bugs & Crash, Test, Check, Read, Window, Add-page, Tab, Error, File, Remove, Work, Android, Flake, Support, Update \\
\bottomrule
\end{tabular}
\end{table*}

\begin{table}[h!]
\centering
\caption{Bug Categories Obtained using NLP}
\label{tab:nlp-categories}
\begin{tabular}{@{}p{4.5cm}p{3.5cm}@{}}
\toprule
\textbf{Category} & \textbf{Benchmark Category} \\
\midrule
Network \& Security Bugs & [Match] \\
Memory Leaks &  \\
Rendering Bugs &  \\
Compatibility \& Web Standards & [Match] \\
Performance Bugs & [Match] \\
Extension Conflicts & [Match] \\
JavaScript Engine Bugs &  \\
User Interface Bugs & [Match] \\
Privacy \& User Data & [Match] \\
\bottomrule
\end{tabular}
\end{table}

%==================== RESULTS ==================%
\section{Results}
\label{result}
%Our findings suggest that embedding domain-specific knowledge into LLMs significantly improves the accuracy \todo{What's the accuracy (F1) score?} of bug classification, as it allows the model to leverage contextual information and semantic similarities between bug descriptions and categories. 
%By applying this approach, bugs can streamline the bug prioritization process, providing a more efficient and scalable solution for browser developers. 
We categorized bugs in Firefox and Chromium using GPT-4's knowledge embedding technique. 
We obtained bench-mark bug-categories to compare with manually by analyzing the browser documentations~\cite{bugDocumentationFirefox,bugDocumentationChromium} as shown in Table~\ref{tab:predefined-categories}. 

\subsection{RQ1: How does LLM identify common bugs in modern web browsers? How does NLP technique perform in contrast?}

To obtain bug categories using GPT-4.o, we embedded knowledge into the LLM from the browser documentations, bug reports, and bug-fix commits. 
The embedding process transformed each document in CKS into a high-dimensional vector enabling similarity-based comparisons.
%To support categorization, we employed a predefined set of bug categories, each with a descriptive label outlining its scope and characteristics. We embedded these category descriptions in the same vector space, allowing for direct comparison with bug report embeddings. Categorization was achieved by calculating cosine similarity between each bug’s embedding and each category’s embedding, assigning each bug to the category with the highest similarity score. This approach provided a structured, similarity-based method for matching bugs to the most semantically relevant categories.
The vectors were compared with category-specific embeddings to classify bugs into categories like ''Networking \& Security'' ''Performance \& Resource Management'' and ''UI/UX \& Accessibility. 

We used ROUGE-W to validate the accuracy of the classification. 
Each category is defined by a description that encapsulates a specific issue type, allowing it to capture its unique semantic meaning. 
This setup enables direct comparison between the documents in CKS, yielding precise accuracy.

For each bug we calculated the cosine similarity between the bug description and the category identified in bug reports.
The category with the highest cosine similarity score is selected as the predicted category, as it reflects the most semantically similar category to the document. 
ROUGE-W calculates a weighted longest common subsequence (LCS) between the predicted category description and the reference (true) category description obtained from the bug reports.

For instance, given a document ``\textit{Clear up old dom perf}'', if the predicted category is ``Developer Tools \& Debugging'', and this matches the reference category, ROUGE-W evaluates term alignment, capturing key phrases ``Developer'', ``Tools'' and ``Debugging''. 
%The resulting ROUGE-W F1 score provides a robust accuracy metric, reflecting how well the model's prediction aligns with the intended category contextually and structurally.
The predicted category is: “Developer Tools \& Debugging,” and the reference category obtained from bug reports is: “Debugging and Developer Tools''.

The LCS between ``\textit{Developer Tools \& Debugging}'' and ``\textit{Debugging and Developer Tools}'' includes the terms ``\textbf{Developer Tools}'' and ``\textbf{Debugging}''. 
This LCS yields a match length of 2 using a weight function \( f(x) = x^2 \):
\[
\text{Weighted LCS} = f(2) = 2^2 = 4
\]
where, $x$ denotes the LCS length.
The precision and recall are calculated based on the following equations.

\[
\text{Precision} = \sqrt{\frac{\text{Weighted LCS}}{\text{Length of Prediction}}} = \sqrt{\frac{4}{3}} = 1.15
\]

\[
\text{Recall} = \sqrt{\frac{\text{Weighted LCS}}{\textit{Length of Reference}}} = \sqrt{\frac{4}{4}} \approx 1.0
\]
where the the prediction length (\textit{"Developer Tools \& Debugging"}) was 3 words, and the reference length (``\textit{Debugging and Developer Tools}'') is 4 words.
The combined F1 score was obtained as:
\[
\text{F1} = 2 \times \frac{\text{ROUGE-W Recall} \times \text{ROUGE-W Precision}}{\text{ROUGE-W Recall} + \text{ROUGE-W Precision}}
\]
Therefore, for this example, our F1 score is:
\begin{equation}
\text{ROUGE-W F1} = 2 \times \frac{1.15 \times 1.0}{1.15 + 1.0} \approx 1.0
\end{equation}

On the other hand, using the traditional NLP approach, we applied pre-processing techniques such as tokenization, stemming, and lemmatization to the bug summary texts from the bug reports and commit messages from revision history.
We used TF-IDF to evaluate the significance of terms within the documents. 
This approach enabled us to identify recurring patterns and relationships in the bug reports, particularly in terms of similar bugs across common browser components.

To gain a deeper understanding of the categories identified by TF-IDF, we applied thematic analysis~\cite{braun2012thematic} on the tokens.
The theme analysis repeating four times by two people led us to the identification of common types of bugs in modern web browsers.
These two individuals also acted as raters for measuring inter-rater reliability of the bug categories.
We used Cohen's Kappa coefficient (\textit{k})~\cite{cohen1960coefficient} to measure the inter-rater reliability to gain confidence on the bug types.
Cohen's Kappa was calculated as:
$$k = \frac{p_o - p_e}{1-p_e}$$ 
where $p_o$ is the observed agreement and $p_e$ is the expected agreement on a bug type by chance.
We calculated $p_o = 0.849$ and $p_e = 0.1016$ that gave us the Cohen's Kappa $k = 0.83$ which falls under the ``Almost Perfect'' level of agreement according to the Cohen's Kappa levels~\cite{mchugh2012interrater}. 
The Cohen's Kappa table is available in our replication package.
The bug categories obtained using NLP are listed in Table~\ref{tab:nlp-categories}.

LLM-based method successfully matched all 10 of the bench-mark categories. In contrast the NLP method, based on cosine similarity, was able to match 6 out of 10 categories effectively (Table~\ref{tab:nlp-categories}) with the benchmark categories (Table~\ref{tab:predefined-categories}).
Our contextual knowledge embedded model GPT-4.o classifies with the F1 score of 94.63\%, outperforming the NLP-based method by 30\% in categorizing bugs while NLP had an F1 score of 64.01\%.
\mybox{gray}{\textbf{Answer to RQ1:} LLM based categorization outperformed NLP based approach with an accuracy (F1-Score) of 94.63\%.}.

\subsection{RQ2: What are the common defects in modern web browsers i.e. Firefox and Chromium?}
% \todo{Discuss what bug categories found using NLP and what categories we found using GPT-4.o. What are the common categories. Then report how many bugs belong to what category. Which category has the highest number of bugs and which category has the least number of bugs.}
%To identify the common defect categories in both web browsers, we analyzed bug reports using two methods, traditional NLP techniques, and LLM (GPT-4.o). 
Through knowledge embedding in GPT-4.o, we identified, categorized, and quantified the types of bugs found in Firefox and Chromium browsers, highlighting areas that might need improvement.

%\begin{figure}
%  \centering
%  \includegraphics[width=1\columnwidth]{predicted-cat-ch.pdf}
%  \caption{LLM-Predicted Categories in Chromium.}
%  \label{fig:predicted-categories-ch}
%\end{figure}
%\begin{figure}
%  \centering
%  \includegraphics[width=1\columnwidth]{predicted-cat-fx.pdf}
%  \caption{LLM-Predicted Categories in Firefox.}
%  \label{fig:predicted-categories-fx}
%\end{figure}
%

%Using NLP we identified categories that include ``User Interface Bugs'', System Stability and Compatibility Bugs, Display and Window Management Bugs, and Cross-Platform Build and Test Reliability Bugs. 
These categories we found through NLP process (Table~\ref{tab:nlp-categories}) focus on issues that could impact user experience and stability across different devices and operating systems, as well as issues related to how browsers handle various visual and functional elements.
%In contrast, LLM based categorization provided a more granular and extensive breakdowns of bug types. 
In contrast, the categories identified through GPT-4.o have a wider coverage (Table~\ref{tab:predefined-categories}) that adds ``File Handling \& System Interaction'', ``Installation, Updates, \& User Preferences'', ``Developer Tools \& Debugging'', and ``Session Management \& Synchronization''. 
This set of categories covers a broader spectrum of technical and user-centered issues, shedding light on various aspects of browser performance, stability, and usability."

\subsubsection{Common Categories and Number of Bugs}
By comparing the number of bugs across bug-categories in Firefox and Chromium %as shown in Figure~\ref{fig:predicted-categories-ch} and Figure~\ref{fig:predicted-categories-fx}, 
we observed notable patterns. 
In Chromium, the ``Developer Tools \& Debugging'', ``UI/UX \& Accessibility'', and ``File Handling \& System Interaction'' were the top most categories of bugs with 39K, 34K, and 13K bugs respectively.
The category with the fewest bugs was ``Networking \& Security'', with only 644 reported issues.

In Firefox, the ``Developer Tools \& Debugging'' category was also the most prevalent, with 96K bugs, followed by ``File Handling \& System Interaction'' (77K) and ``UI/UX \& Accessibility'' (50K) as the second and third most bug-prone common categories respectively.
Similar to Chromium, ``Networking \& Security'' was the least frequent bug-category with only 4.9K bug reports.

The numbers show that both Firefox and Chromium struggle with ``Developer Tools \& Debugging'', and ``UI/UX \& Accessibility'' types of bugs which suggests that these are the challenging areas where more attention is needed. 
Improving these areas could greatly enhance both the developer experience and the overall usability for end-users.
On the other hand, ``Networking \& Security'' bugs are the fewest in both browsers indicating that they may handle security concerns relatively well compared to other areas.
\mybox{gray}{\textbf{Answer to RQ2: } Paying a deeper attention to debugging-tools, user interface, and accessibility related features could help both Firefox and Chromium address some of their most persistent issues. 
This provides a clear direction for future work, with the potential to significantly enhance browser stability, usability, and user satisfaction.}

\subsection{RQ3: Do highly effort-consuming bugs exist, and if so, are they prevalent in both browsers?}
Firefox experienced 2.5 times more bugs than Chromium in its lifetime.
We identified a total of 370K bugs in Firefox while Chromium has only 143K. 
In adherence to coding conventions outlined in both Firefox~\cite{firefoxcheckinrules} and Chromium~\cite{chromium-commit-rule}, when a commit is linked to a specific bug, the bug ID is incorporated into the commit message. We considered these commits as bug-fix commits as they are presumed to attempt a fix for the bug. 

\begin{figure}[!ht]
  \centering
  \includegraphics[width=0.45\textwidth, height=0.35\textheight]{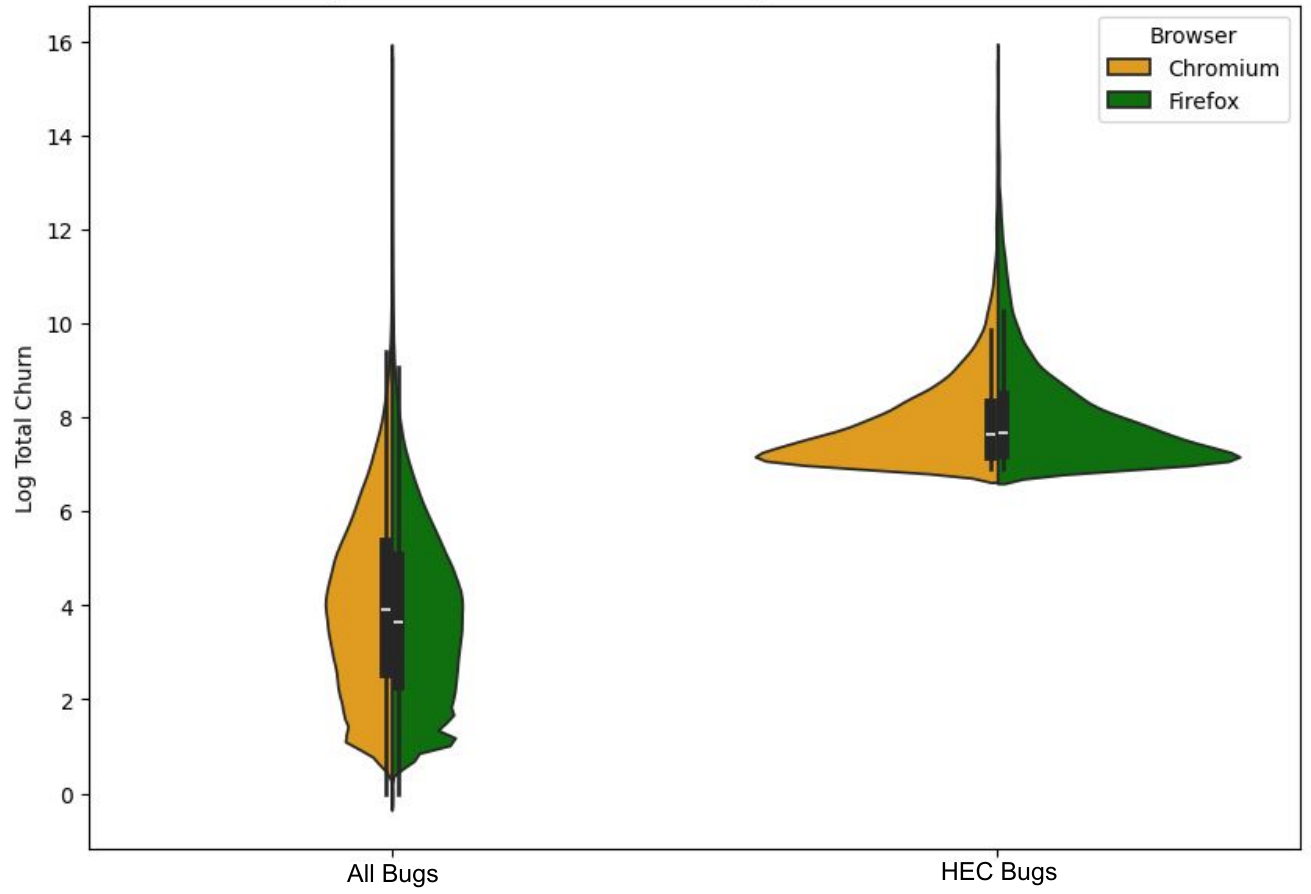}
  \caption{Bug-fix Effort in Chromium and Firefox}
  \label{fig:churn-per-bug-ch-fx}
\end{figure}

We considered the number of code changes (i.e. churn) as effort. 
The violin plots in Figure~\ref{fig:churn-per-bug-ch-fx} represent the bug-fix efforts in Firefox and Chromium. On average, each bug in Firefox requires 8.0 churns, with a maximum of 6.2M 
Bugs in Chromium require almost similar effort (7.0) with a maximum of 4.6M churns.

To determine Highly Effort Consuming (HEC) bugs, we considered 1K churns as a threshold.
We found HEC bugs in both web browsers.
The number of HEC bugs is also higher in Firefox than in Chromium.
There are \textbf{18,625} HEC bugs in Firefox while Chromium has \textbf{8,855} HEC bugs. 
Table~\ref{tab:top-effort-consuming-bugs} shows the top 5 high effort consuming bugs in both browsers sorted by number of churns. 
Most of the HEC bugs in Chromium occurred in the Third Party code while
Firefox's top HEC bugs are not clustered in one or two components.
\mybox{gray}{\textbf{Answer to RQ3:} Although, median churn per HEC bug is very similar, there are 18K and 8.8K HEC bugs in Firefox and Chromium respectively. Most of the top HEC bugs in Chromium are found in Third Party components.
}

\begin{table}[htbp]
\caption{Top 5 Highly Effort Consuming (HEC) Bugs.}
\label{tab:top-effort-consuming-bugs}
\centering
\begin{tabular}{@{}lll@{}}
\toprule
\multicolumn{3}{c}{\textbf{Firefox}}\\
\midrule
\textbf{Bug ID} & \textbf{Component} & \textbf{Churn}\\
1075758 & ICU & 6.2M\\
1773223 & Media & 4.4M\\
1376873 & Third Party & 3.3M\\
1654112 & Third Party & 3.2M\\
1410245 & ICU & 3.0M\\
%1396158 & Pocketsphinx & Media & 133449\\
%\hline
%1051146 & English Model & Media & 133437\\
%\hline
%1478354 & WPT Manifest Update & Web Platform & 112136\\
%\hline
%1338567 & Debugger Test & Devtools & 112138*\\
%\hline
%1439839 & Babel Types Ref. & Devtools & 92490\\
%\hline
\midrule
\multicolumn{3}{c}{\textbf{Chromium}}\\
\midrule
\textbf{Bug ID} & \textbf{Component} & \textbf{Churn} \\
14756 & Third Party & 4.6M\\
340757 & Third Party & 2.1M\\
701518 & Third Party & 1.8M\\
579743 & Third Party & 1.6M\\
238755 & Third Party & 1.3M\\
%665328 & WebGL & GPU & 7843\\
%\hline
%923083 & Histogram Eraser & Tools & 5853\\
%\hline
%1152197 & WebGL Update Bot & GPU & 5537\\
%\hline
%1028298 & WebGL Runtimes & GPU & 5517\\
%\hline
%1030096 & Histogram Expire & Tools & 5154\\
%\hline
\bottomrule
\end{tabular}
\end{table}

%\todo{What type of bugs are these HEC bugs}

%\begin{figure}
%  \centering
%  \includegraphics[width=1.0\columnwidth]{component-bugs-ch-fx.pdf}
%  \caption{Bugs in Components for Chromium and Firefox}
%  \label{fig:common-component-bugs}
%\end{figure}
%
\begin{figure*}[htbp]
  \centering
  \resizebox{\textwidth}{!}{\includegraphics{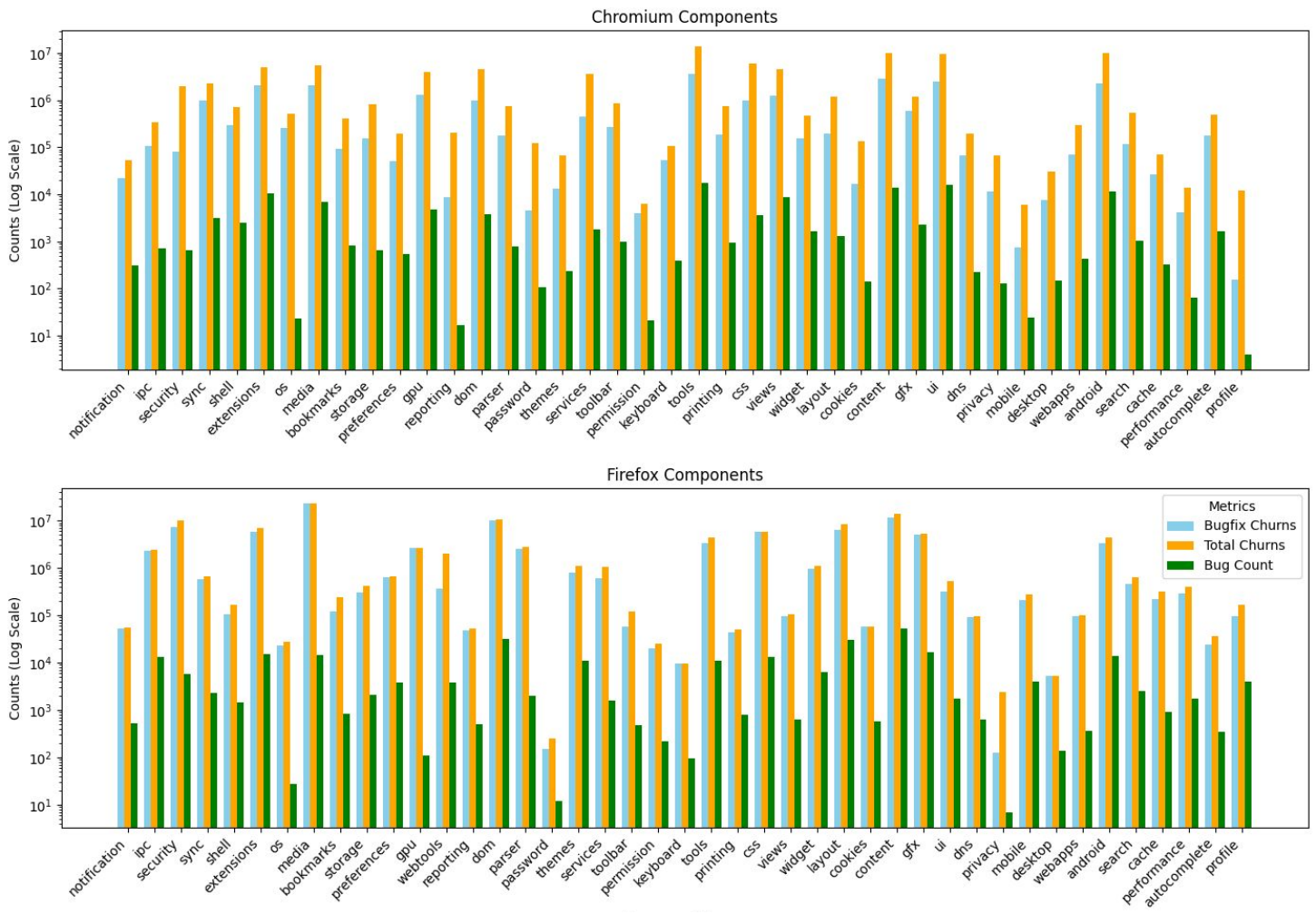}}
  \caption{Bugs in Components for Chromium and Firefox}
  \label{fig:common-component-bugs-bar}
\end{figure*}

\subsection{RQ4: What are the common components in modern web browsers, and which of them need more attention from developers?}

It is important to know which browser components are affected by more bugs, i) to make an informed decision when developers are prioritizing their test cases, ii) allocating maintenance effort, and integrating new features or adding new component into the system.

There are 43 components that are common in both Firefox and Chromium. 
As we see in Table~\ref{tab:top-defect-prone-components}, Content, DOM, Layout, GFX (Graphics), Extensions, Media, Android, IPC, CSS, and Tools are the top ten components in Firefox that experience the more than 10K bugs. 
On the other hand, Tools, UI, Content, Android, Extensions are the five components that are highly defect-prone with number of bugs greater than 10K.
In the median case the 43 common components had 1744 bugs in Firefox, and 811 bugs in Chromium.

Among the top ten defect-prone components in these two browsers five (DOM, Content, Extensions, Media, and Android) are common in both, marked with an ``*''.
%Figure~\ref{fig:common-component-bugs} represents the violin plots comparing bugs in Firefox and Chromium components. 
In general components in Firefox experience significantly high number of bugs compared to Chromium. 
In median case (regardless of common/not-common) Firefox components experienced 10.6K bugs while Chromium had only 2K.
This further explained in Figure~\ref{fig:common-component-bugs-bar} representing the bugs in the common components in the two browsers where X-axis shows the component names and Y-axis presents the Number of Bugs (green), Bug-Fix Churns (cyan), and Total Churn (yellow) per component.

\begin{table}[htbp]
\caption{Highly Defect-Prone Components (10K+ Bugs).}
\begin{center}
\begin{tabular}{@{}lrr@{}}
\toprule
\multicolumn{3}{c}{\textbf{Firefox}}\\
\midrule
\textbf{Component} & \textbf{Number of Bugs} & \textbf{Bug-Fix Churns}\\
Content* & 53K & 11.9M\\
DOM* & 31K & 9.9M\\
Layout & 30K & 6.5M\\
GFX & 16K & 4.9M\\
Extensions* & 15K & 5.9M\\
Media* & 14.2K & 22.9M\\
Android* & 13.8K & 3.4M\\
IPC & 13.1K & 2.3M\\
CSS & 13K & 5.8M\\
Tools & 11K & 3.3M\\
\midrule
\multicolumn{3}{c}{\textbf{Chromium}}\\
\midrule
Tools & 17.5K & 3.7M\\
UI & 16.2K & 2.5M\\
Content* & 14.0K & 2.8M\\
Android* & 11.7K & 2.3M\\
Extensions* & 10.4K & 2.0M\\
Views & 8.5K & 1.2M\\
Media* & 6.9K & 2.0M\\
GPU & 4.8K & 1.3M\\
DOM* & 3.7K & 1M\\
CSS & 3.6K & 0.9M\\
\bottomrule
\end{tabular}
\end{center}
\end{table}

Most of the components in Firefox experience high number of bugs as well as high number of bug-fix commits compared to the components in Chromium.
In median case Firefox components receive 304K bug-fix churns while Chromium components receive around a half (155K) of that.
In Figure~\ref{fig:common-component-bugs-bar}, although, the trends look pretty similar in these two browsers, there are some surprising odds.
For example, ``Profile'' in Chromium receives the least number of bugs and bug-fix churns, in Firefox it is one of the high defect-prone components. 

The top-10 bug-fix effort (churns) does not strongly correlate to the number of bugs.
A Pearson correlation between the bugs and the bug-fix efforts in Firefox and Chromium gave us coefficients of 0.27 and 0.48 respectively indicating a very weak correlation between number of bugs and bug-fix churns for these highly defect-prone components.
For all components regardless of top-10, we observed correlation coefficients in Firefox and Chrome as 0.67 and 0.65 respectively.
On the other hand, for all components excluding the top-10, we observed the correlation coefficients in Firefox and Chrome as 0.48 and 0.76 respectively.
\mybox{gray}{\textbf{Answer to RQ4: } There are 43 common components in Firefox and Chromium. Firefox components are highly defect-prone and effort consuming. Especially, DOM, Content, Extensions, Media, and Android components are the commons of the top-10 defect-prone components in both of these modern browsers.
}

\section{Threats to Validity}
\label{threats}
\subsubsection{Internal validity}
% Component and sub-components 
We conducted a manual investigation to validate the component names, cross-referencing with Firefox, and Chromium documentations for accuracy. 
Furthermore, by cross-validating against bug reports, we identified additional components, concluding in a unified, distinct set.

% Bugfix commits
We retrieved bug-fix commits from the commit history using a regular expression that searches for a pattern, which conforms to the common standard practiced in both Chromium~\cite{chromium-commit-rule} and Firefox~\cite{firefoxcheckinrules}.
However, there might be instances where the standard wasn’t followed by the developers. 
Our regular expression-based approach would not capture those commits.
To validate this, we checked how many bug-fix commits our regular expression could map with the bugs from the bug reports. 
We found 96\% bugs in the bug reports were retrieved from the commits. 
Given the high accuracy of our regex method, we decided to proceed with this dataset.

Labeling the code revision data with components was challenging because components often overlapped at the directory level.
Some of the labels may not be fully accurate. 
However, we had to stick with a common strategy.
We applied random-sampling~\cite{olken1993random} technique to assure the performance of our strategy. 

%Cluster labeling
In cluster analysis, we filtered out irrelevant and overly common words by setting up thresholds.
Finding an appropriate threshold is tricky. 
A marginally different threshold would give us slightly different results.
To validate this threat, we calculated the TF-IDF difference for each word between documents to measure relevance and used GTSS~\cite{ZHANG2020540} to derive asymmetric thresholds for word extraction, which is one of the most effective approaches for determining TF-IDF thresholds.

%Since it is challenging to ensure perfect classification when dealing with cross-cutting concerns like Security, which can impact multiple components such as Networking and Extensions.
%To mitigate this, we relied on documented best practices for component assignment and cross-referenced historical bug reports to improve consistency. In the cases where a bug had significant overlap across multiple areas, we attempted to follow prior labeling conventions used in Chromium and Firefox issue tracking. 
%That said, labeling strategies continue to evolve, and further refinements—such as introducing multi-component tagging or hierarchical classifications—could enhance accuracy in the future.

To avoid bias in GPT-4.0, we filtered out words that were too common or didn’t add much value to the bug categorization. 
We did this based on the frequency of terms and ensured only the most relevant ones were considered. 
Additionally, we double-checked the results by comparing them with known bug data and manually labeling some to make sure the model wasn’t just picking up on frequently used terms without real meaning. 
We also used a random sampling technique to variate terms, which helped prevent the model from overfitting to certain words.

\subsubsection{External validity}
%This study focuses on two open-source web browsers that share similarities in terms of their life span, release structure, release frequency, and architecture.
The two web browsers in this study are widely popular and representative of modern web browsers.
We identified browser components based on our own approach, primarily from the revision data. 
We used browser documentations to filter out many false positives. 
This method can be applicable to any web browser if we have access to the source code. 
%This will be useful for readers, practitioners, and future researchers to know what are the components we have identified based on our strategy leaving the ground of criticism and expansion of the study open.

%Cluster labeling 
%Categorizing bugs presents potential for wider application beyond Firefox and Chromium, thereby enhancing the external validity of our bug commonality analysis. 
%LLM predicted categories may serve as a foundation for classifying and prioritizing issues in modern web browsers. 

\section{Conclusion and Future Work}
\label{conclusion}
This study presents a novel investigation of the common bugs in modern web browsers. 
By leveraging contextual knowledge from the browser documentations, commit messages, and bug descriptions, we utilized GPT-4.0 to analyze and categorize bugs effectively, comparing its performance against the traditional NLP approach of TF-IDF. 
Our findings shed light on defect-prone components in modern web browsers, providing actionable insights for the developers
%to prioritize tasks and make informed decisions when addressing browser defects. 
%These contributions not only advance our understanding of browser-related bugs but also offer practical value to the software engineering community, particularly to modern browser developers.

% Cluster and Categories
%Bugs can be clustered in three groups in Firefox and two for Chromium. 
Firefox and Chromium mostly suffer from ``Developer Tools'', ``Debugging'', ``UI/UX Accessibility'', and ``File Handling'' types of bugs. 
It seems both browsers are more careful about Network and Security issues since they are the least frequent in both Firefox and Chromium. 
%
%The thematic analysis in the NLP based approach explored that both browsers experience similar Functionality and Compatibility issues, covering security vulnerabilities, memory leaks, rendering problems, web standards compatibility, performance issues, extension conflicts, JavaScript Engine and UI glitches, network issues, privacy concerns, and plugin vulnerabilities.
% Effort consuming bugs
Both browsers have extremely high effort consuming (HEC) bugs. 
However, most of the HEC bugs in Chromium are coming from ``Third Party'' code.
Majority of components in Firefox are highly defect-prone compared to those in Chromium.
% Defect prone components
%Almost 50\% of the components are common in the two browsers. 
%However, they are not equally defect-prone. 
Firefox has twice as many highly defect-prone components as Chromium; however, five of these components are common to both browsers.
These findings will allow developers to take informed decisions before modifying such components. 
The developers of other web applications will also benefit from being aware of the issues of the browser components.%, additional safety measure before making any design decision.
%Developers will also be able to prioritize testing and bug-fixing effort for highly defect-prone components.
%This study highlights the performance of LLM based classification over traditional NLP methods leveraging deep contextual understanding, in categorizing complex bug descriptions. 
%It provides insights into common defects, defect-prone components, and the developer effort involved in modern web browsers.

%We aim to implement topic modeling techniques to automatically categorize and bug reports into topics/themes.
%We also aim to conduct future studies for security assessments, code-smells and architectural violation analysis for performance optimization, and privacy concerns for better development of modern web browsers.
%Sentiment analysis on bug reports and user feedback would gauge the emotional tone and user satisfaction related to specific defects. 
Our study opens several promising avenues for future research. First, incorporating temporal analysis to study how bug trends evolve over time could help identify emerging challenges and shifting priorities in browser development. 
Second, exploring the integration of automated bug triaging systems powered by large language models (LLMs) could streamline bug classification and prioritization for developers. 
Additionally, investigating the root causes of browser bugs and their impacts across different components as well as browser-based applications and analyzing their correlation with design patterns could offer deeper insights into improving browser reliability. %Finally, applying similar methodologies to other complex software ecosystems, such as operating systems, cloud platforms, could generalize the findings and advance software quality assurance.
Furthermore, Firefox and Chromium have recently shifted to a shorter release cycle of 4 weeks from 6 weeks.
Future studies may investigate the trends in bugs and the impact of different lengths of release cycles.
%Furthermore, since bug reports and bug-fix commits contain valuable detailed information about bugs, impacts, and locations at a very low level, Large Language Models could be utilized for advanced analysis and automated defect prediction.

%\appendix

\printbibliography

\end{document}